\documentclass{article}
\usepackage{graphicx}
\usepackage{txfonts}
\usepackage{natbib}
\usepackage{hyperref}
\usepackage{booktabs}
\usepackage[switch]{lineno}
\nolinenumbers

\usepackage{amsmath}

\begin{document}

\title{A linear residual structure across galaxy rotation curves }

\author{Hosik Lee \\ School of Energy and Chemical Engineering,
  Department of Energy Engineering, \\ Ulsan National Institute of
  Science and Technology (UNIST), \\Ulsan 44919, Republic of
  Korea\\ \texttt{hslee@unist.ac.kr} }

\maketitle

\begin{abstract} Galaxy rotation curves exhibit systematic discrepancies
  between the observed dynamics and the gravitational contribution
  expected from baryonic matter. Identifying empirical regularities in
  these discrepancies may provide insight into the organization of
  galaxy dynamics.  We investigate whether the residual component of
  galaxy rotation curves contains a common structure across galaxies
  spanning a broad range of masses and morphologies.  Using
  rotation-curve data from the SPARC and LITTLE THINGS surveys, we
  analyze residual velocity-squared profiles after accounting for the
  baryonic contribution and allowing for uncertainties in the baryonic
  normalization.  We find that the residuals are not randomly
  distributed but instead follow a common linear pattern across a
  diverse galaxy population.  Population-level analysis shows that the
  data preferentially select this linear residual structure over
  alternative radial dependences.  The residual component separates
  into a mass-coupled contribution and a second contribution that
  remains nearly independent of galaxy mass.  These empirical trends
  are observed across both spiral and dwarf galaxy samples.  The
  existence of a common residual structure across galaxies spanning a
  broad range of masses and morphologies provides a new empirical
  constraint on theories of galaxy dynamics.
\end{abstract}

Galaxy rotation curves, which trace the orbital velocities of stars
and gas as a function of distance from the galactic centre, provide
one of the most direct probes of the distribution of matter in
galaxies. Observations over the past decades have consistently shown
that the observed rotational velocities exceed those expected from the
visible baryonic components (stars and gas) alone \cite{Rubin1980}.In
many galaxies, particularly at large radii, the discrepancy becomes
comparable to or larger than the baryonic contribution itself.  

The physical origin of this residual component remains one of the
central questions in astrophysics. Within the standard cosmological
framework, it is commonly attributed to dark matter halos surrounding
galaxies \cite{Navarro1996,Navarro1997}. Alternative explanations
based on modifications of gravitational dynamics have also been
extensively explored \cite{Milgrom1983,Sanders2002}. Despite their
different physical interpretations, both approaches seek to explain
the same observational phenomenon: the residual component that remains
after accounting for baryonic matter.

Here we investigate the residual component of galaxy rotation curves
using the SPARC sample \cite{Lelli2016} and the LITTLE THINGS sample
\cite{Hunter2012,Oh2015}. Rather than interpreting the residuals
within a specific theoretical framework, we ask a simpler empirical
question: do galaxy rotation-curve residuals exhibit an underlying
organizational structure?

We find that the residuals are not randomly distributed. Instead, they
exhibit a remarkably simple and organized structure across a diverse
population of galaxies spanning a wide range of masses, morphologies,
and dynamical properties. Population-level analysis shows that the
data favour a common linear residual pattern over alternative
models. Furthermore, the residual structure separates into two
distinct components: one that scales systematically with baryonic mass
and another that remains nearly independent of galaxy mass. These
results reveal an unexpected degree of organization in galaxy
rotation-curve residuals and provide a new empirical perspective on
the relationship between baryonic matter and galaxy dynamics.

\begin{figure*}[!t]
\centering
\includegraphics[width=\textwidth]{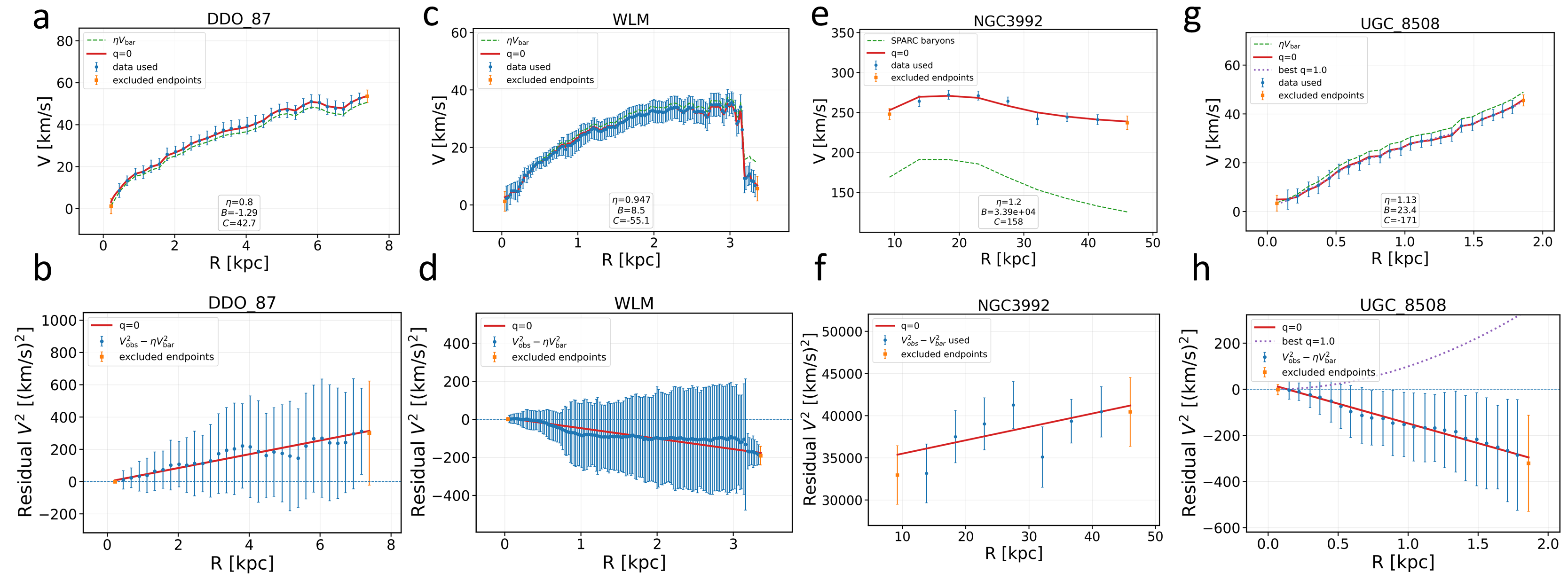}
\caption{\textbf{A linear residual structure in galaxy rotation
    curves.}  Representative galaxies from the SPARC and LITTLE THINGS
  samples. Upper panels show the observed rotation curves (blue
  points), baryonic contributions (green lines), and best-fit models
  (red lines). Lower panels show the residuals, $R(r)=V_{\rm
    obs}^{2}(r)-\eta V_{\rm bar}^{2}(r)$, together with the best-fit
  linear models, $R(r)=B+Cr$.  }
\end{figure*}

For circular orbits, the balance between centrifugal and gravitational
acceleration ($V^2/r=g(r)$) motivates the use of velocity-squared
quantities as tracers of the underlying gravitational contribution.

\begin{equation}
R(r)=V_{\rm obs}^{2}(r)-\eta V_{\rm bar}^{2}(r).
\end{equation}

 where $V_{\rm obs} (r)$ is the observed rotation curve, $V_{\rm
   bar}(r)$ represents the contribution from stars and gas. The
 normalization factor $\eta$ accounts for uncertainties in the
 inferred baryonic mass, primarily associated with stellar
 mass-to-light ratios and gas content \cite{Lelli2016}.

To investigate whether the residual structure is genuinely linear, we
fitted the residuals using the generalized model

\begin{equation}
R(r)=B+Cr^{q+1},
\end{equation}

where $q$ is treated as a free parameter.  The search was restricted
to the range $0\le q\le1$, spanning radial dependences from linear
($q=0$) to quadratic ($q=1$). Values near $q=-1$ were excluded because
the resulting term becomes degenerate with the constant component $B$.
The preferred value of $q$ was then determined from the data as shown
by Figure~1.

Figure~1 presents four representative galaxies spanning a wide range
of galaxy properties. DDO~87 (Fig.~1a,b) provides one of the simplest
examples in the sample, showing a smooth rotation curve and a clear
linear residual pattern. WLM (Fig.~1c,d) represents a much more
irregular system with substantial observational scatter, yet the
residuals remain approximately linear. NGC~3992 (Fig.~1e,f), a massive
spiral galaxy, exhibits one of the largest residual amplitudes in the
sample, but again follows the same linear trend. Finally, UGC~8508
(Fig.~1g,h) illustrates a case in which several alternative models
provide similarly good fits. Even in this situation, the residuals are
naturally described by the linear form $R(r)=B+Cr$.  Taken together,
these examples suggest that the linear residual pattern is not a
feature of any particular galaxy type, but a common property shared
across a diverse galaxy population.

The emergence of the same residual structure across such diverse
systems suggests that the residual component is not dominated by
random scatter, but instead follows an organized radial pattern.

\begin{figure}[t]
\centering
\includegraphics[width=\columnwidth]{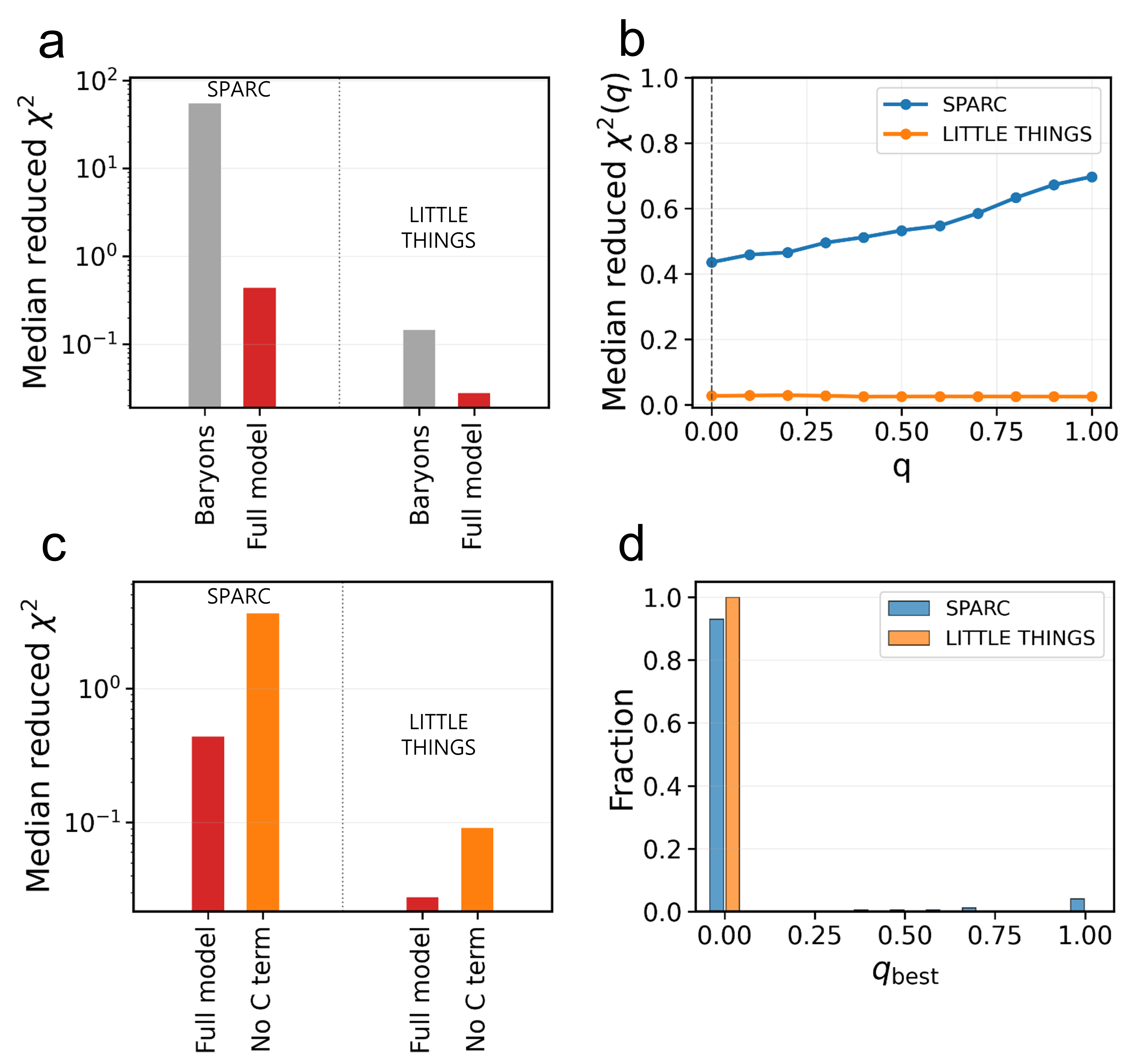}
\caption{\textbf{Population-level selection of the linear residual model.}
\textbf{a,} Distribution of reduced $\chi^{2}$ values for baryon-only and full residual models.
\textbf{b,} Median reduced $\chi^{2}$ as a function of the residual exponent $q$ in the generalized model $R(r)=B+Cr^{q+1}$.
\textbf{c,} Distribution of reduced $\chi^{2}$ values for the full model and models without the linear residual term.
\textbf{d,} Distribution of best-fit values of $q$.
Both the SPARC and LITTLE THINGS samples independently favour $q\approx0$.
}
\end{figure}

Figure~2 demonstrates that the linear residual structure is favoured
across the full SPARC and LITTLE THINGS samples. To quantify the
agreement between models and observations, we use the reduced
$\chi^2$, a statistical measure of goodness-of-fit for which values
near unity generally indicate consistency with the data, whereas
substantially larger values indicate poor agreement.

Baryon-only models yield median reduced $\chi^2$ values of
approximately 55 in the SPARC sample (Fig.~2a), indicating that the
observed rotation curves cannot be reproduced by the baryonic
contribution alone. Including the residual component dramatically
improves the agreement with the observations, reducing the median
reduced $\chi^2$ by more than two orders of magnitude.

We next test whether the residual structure is genuinely linear by
fitting the generalized model $R(r)=B+Cr^{q+1}$ while allowing $q$ to
vary between 0 and 1. For the SPARC sample, the best overall agreement
is obtained near $q=0$ (Fig.~2b), indicating that the data
preferentially select a linear residual form $R(r)=B+Cr$. In the
LITTLE THINGS sample, the dependence of fit quality on $q$ is
weaker. Nevertheless, representative systems such as UGC~8508
(Fig.~1h) recover an approximately linear residual relation when $q=0$
is adopted, consistent with the common residual structure observed
across the broader galaxy population.

Removing the linear residual term leads to a marked deterioration in
fit quality (Fig.~2c), demonstrating that the improvement cannot be
attributed simply to additional free parameters. Consistently, the
distribution of best-fit values is strongly concentrated near
$q\simeq0$ (Fig.~2d), indicating that the linear residual structure
emerges directly from the observations rather than being imposed by
the model.

Figure~3 examines how the coefficients $B$ and $C$ vary across the
galaxy population. Three distinct branches are apparent in the
distribution of $B$ (Fig.~3a): a positive-$B$ branch, a negative-$B$
branch, and a near-zero branch ($|B|<20$ )centred around
$B\approx0$. The positive-$B$ branch follows an approximate scaling
relation

\begin{equation}
B \propto M_{\rm bar}^{0.65},
\end{equation}

whereas the negative-$B$ branch follows a shallower relation,

\begin{equation}
|B| \propto M_{\rm bar}^{0.41}.
\end{equation}

Galaxies with $B\approx0$ are particularly common among dwarf systems in the LITTLE THINGS sample.

In contrast, the coefficient $C$ exhibits little dependence on baryonic mass (Fig.~3b). A power-law fit yields

\begin{equation}
|C| \propto M_{\rm bar}^{0.023\pm0.046},
\end{equation}

which is statistically consistent with no mass dependence. The linear
residual structure therefore separates into a mass-coupled component,
represented by $B$, and a nearly mass-independent component,
represented by $C$.

\begin{figure}[t]
\centering
\includegraphics[width=\columnwidth]{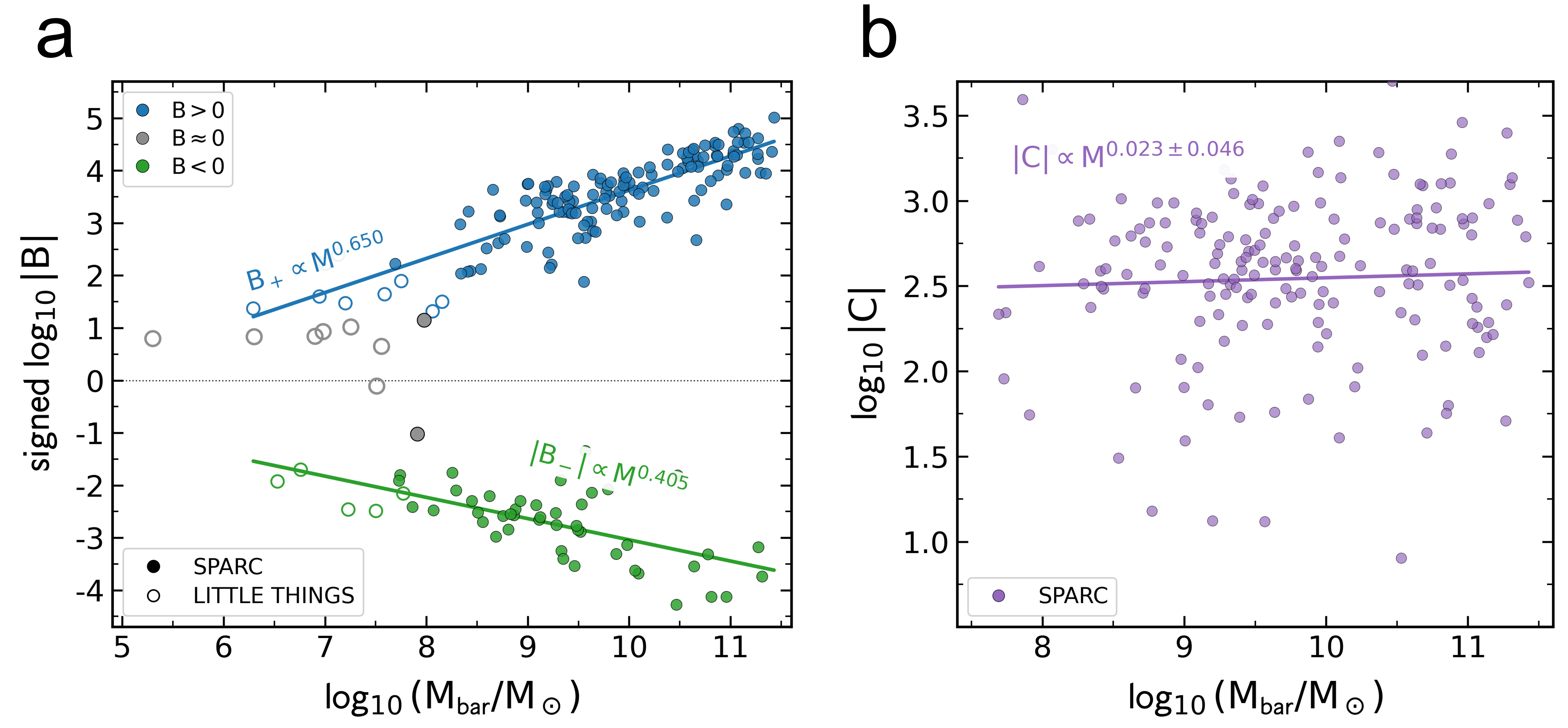}
\caption{\textbf{Mass-coupled and mass-independent residual components.}
\textbf{a,} Dependence of the residual coefficient $B$ on total baryonic mass. Positive-$B$, negative-$B$, and $B\approx0$ branches are present. Dashed lines show power-law fits to the positive-$B$ and negative-$B$ populations.
\textbf{b,} Dependence of the residual coefficient $C$ on total baryonic mass. The best-fit relation,
$|C|\propto M_{\rm bar}^{0.023\pm0.046}$,
indicates little dependence on galaxy mass over the full sample.
}

\end{figure}

The existence of a common residual structure is noteworthy because the
galaxies considered here span a broad range of masses, morphologies,
and rotation-curve shapes. The preference for a linear residual
relation is observed both in individual systems and at the population
level. This suggests that the residual component contains information
that is not captured by baryonic mass models alone and may represent
an additional level of organization within galaxy dynamics.

A second notable result is the separation of the residual structure
into two distinct components. The coefficient $B$ exhibits clear mass
dependence and separates into positive, negative, and near-zero
branches, whereas the coefficient $C$ remains nearly independent of
galaxy mass over the full sample. The positive branch follows a
scaling relation close to $B\propto M_{\rm bar}^{0.65}$, which is
approximately consistent with a two-thirds power law. Large positive
values of $B$ are predominantly associated with massive spiral
galaxies, whereas dwarf galaxies are frequently found near the
$B\approx0$ branch. This systematic transition suggests that the
mass-coupled component is linked not only to baryonic mass itself, but
also to the emergence of large-scale galactic structure. The
approximate $M_{\rm bar}^{2/3}$ scaling is particularly intriguing
because it differs from the linear dependence expected for a quantity
that simply traces mass, indicating that the residual structure may be
associated with a more global property of galaxy organization.

Regardless of interpretation, the empirical regularities identified
here provide new observational constraints that any successful theory
of galaxy dynamics should reproduce. Future studies may determine
whether the linear residual structure and its associated scaling
relations emerge naturally within existing frameworks or point toward
a previously unrecognized aspect of galaxy dynamics. In particular,
understanding the origin of the mass-coupled component and its
transition from dwarf to spiral galaxies may offer new insight into
the relationship between baryonic matter and galactic structure.

\section*{Methods}

The SPARC sample consists of 175 late-type galaxies with high-quality
rotation curves and baryonic mass models derived from Spitzer
photometry and HI observations \cite{Lelli2016}. The LITTLE THINGS
sample consists of nearby dwarf irregular galaxies with resolved HI
rotation curves \cite{Hunter2012,Oh2015}.  Twenty-two galaxies
satisfying the adopted data-quality and radial-coverage criteria were
included in the analysis.

For SPARC galaxies, the baryonic contribution was constructed using
the fiducial stellar mass-to-light ratios adopted by the SPARC survey,
namely 0.5 and 0.7 for stellar disks and bulges, respectively
\cite{Lelli2016}. The model was written as

\begin{equation}
\begin{split}
V_{\rm obs}^{2}(r) ={}&
V_{\rm gas}^{2}(r)
+ \eta \left[
0.5\,V_{\rm disk}^{2}(r)
+ 0.7\,V_{\rm bul}^{2}(r)
\right]  \\
&+ B + C r^{q+1}.
\end{split}
\end{equation}

where $\eta$ accounts for uncertainties in the overall baryonic
normalization and was allowed to vary within the range $0.8 \le \eta
\le 1.2$.

For LITTLE THINGS galaxies, the residual component was defined
relative to the published baryonic rotation curve,

\begin{equation}
V_{\rm obs}^{2}(r) = \eta V_{\rm bar}^{2}(r) + B + Cr^{q+1}.
\end{equation}

The signed convention $V^{2}_{\rm signed}=V|V|$ was adopted whenever
negative velocity contributions were present.

The exponent $q$ was scanned over the range $0\le q\le1$ in steps of
0.1.  For each galaxy, the parameters $(\eta,B,C)$ were determined
through $\chi^{2}$ minimization for every value of $q$.  When
sufficient radial points were available, the innermost and outermost
measurements were excluded from the fit in order to reduce
edge-dominated systematics.

To evaluate whether a galaxy preferred a value of $q$ different from
zero, we compared the minimum $\chi^{2}$ obtained over the scanned $q$
range with the corresponding value at $q=0$. Following the standard
likelihood-ratio criterion for one effective degree of freedom,
solutions satisfying

\begin{equation}
\Delta\chi^{2} = \chi^{2}(q=0) - \chi^{2}_{\rm min} \le 2
\end{equation}

were classified as statistically consistent
with $q=0$.

The Supplementary Information contains the full
set of rotation-curve fits, residual profiles, and
best-fit parameters for all 175 SPARC galaxies and
22 LITTLE THINGS galaxies included in this study.

% ------------------------------------------------------------------
\section*{Acknowledgements}

  We gratefully acknowledge the authors of the SPARC and LITTLE THINGS
  surveys for making their data publicly available. Without their
  efforts in constructing, curating, and openly sharing these
  high-quality datasets, this work would not have been possible.

% ------------------------------------------------------------------
\bibliographystyle{aa}
\bibliography{references}

\end{document}